\documentclass[preprints,article,accept,pdftex,moreauthors]{mdpi}

\usepackage{isotope}
\usepackage{booktabs}
\usepackage{longtable}

\usepackage{xcolor}      
\usepackage{soul}        
\sethlcolor{white}        

\firstpage{1}
\pubvolume{1}
\issuenum{1}
\articlenumber{0}
\pubyear{2025}
\copyrightyear{2025}
\datereceived{}
\dateaccepted{}
\datepublished{}
\hreflink{https://doi.org/} 

\Title{
	Robust Method for Confidence Interval Estimation in Outlier-Prone Datasets: Application to Molecular and Biophysical Data
}

\TitleCitation{
	Robust Method for Confidence Interval Estimation in Outlier-Prone Datasets: Application to Molecular and Biophysical Data
	}

\Author{Victor V. Golovko$^{1}$*\orcidA{}}

\AuthorNames{Victor V. Golovko}

\AuthorCitation{Golovko, V.V.;}

\address{%
$^{1}$\quad Canadian Nuclear Laboratories, 286 Plant Road, Chalk River, ON, Canada, K0J~1J0; \\
}

\corres{Correspondence: victor.golovko@cnl.ca (V.V.G.)}

\abstract{
Estimating confidence intervals in small or noisy datasets is a recurring challenge in biomolecular research, particularly when data contain outliers or exhibit high variability. This study introduces a robust statistical method that combines a hybrid bootstrap procedure with Steiner's most frequent value (MFV) approach to estimate confidence intervals without removing outliers or altering the original dataset. The MFV technique identifies the most representative value while minimizing information loss, making it well suited for datasets with limited sample sizes or non-Gaussian distributions.
To demonstrate the method's robustness, we intentionally selected a dataset from outside the biomolecular domain--fast-neutron activation cross-section of the $^{109}\text{Ag}(\text{n},2\text{n})^{108\text{m}}\text{Ag}$ reaction from nuclear physics. This dataset presents large uncertainties, inconsistencies, and known evaluation difficulties. Confidence intervals for the cross-section were determined using a method called the MFV-hybrid parametric bootstrapping (MFV-HPB) framework. In this approach, the original data points were repeatedly resampled, and new values were simulated based on their uncertainties before calculating the MFV. Despite the dataset's complexity, the method yielded a stable MFV estimate of \(709~\mathrm{mb}\), with a 68.27\% confidence interval of \([691, 744]~\mathrm{mb}\), illustrating the method's ability to provide interpretable results in challenging scenarios.
Although the example is from nuclear science, the same statistical issues commonly arise in biomolecular fields, such as enzymatic kinetics, molecular assays, and diagnostic biomarker studies. The MFV-HPB framework provides a reliable and generalizable approach for extracting central estimates and confidence intervals in situations where data are difficult to collect, replicate, or interpret. Its resilience to outliers, independence from distributional assumptions, and compatibility with small-sample scenarios make it particularly valuable in molecular medicine, bioengineering, and biophysics.
}

\keyword{most frequent value; hybrid parametric bootstrapping; robust statistical method; fast-neutron activation cross-section of the $^{109}\text{Ag}(\text{n},2\text{n})^{108\text{m}}\text{Ag}$ reaction; half-life of the $^{108\text{m}}\text{Ag}$ }

\begin{document}

\section{ Introduction }

Reliable statistical analysis is essential in many areas of biomolecular sciences, including bioinformatics~\cite{Jumper.etalHighlyAccurateProtein2021}, biomedical diagnostics~\cite{Ahmadraji.etalBiomedicalDiagnosticsEnabled2017}, and bioprocess engineering~\cite{StaffPuttingBiotechnologyWork1992}. Researchers in these fields often work with complex datasets that are small in size, contain outliers, or deviate from normal (Gaussian) distributions~\cite{DeTorrente.etalShapeGeneExpression2020}. These characteristics can make conventional statistical approaches, such as the arithmetic mean or least squares fitting, less reliable or misleading.

The most frequent value (MFV) method is a robust alternative that estimates the central value of a dataset based on its densest region. This makes the proposed method highly resistant to outliers and better suited for datasets with irregular structures or measurement noise. Because it preserves more of the original information in the data, MFV is particularly valuable in biomolecular contexts, where measurements are often costly or variable, and data points cannot be easily discarded.

In bioinformatics, MFVs can improve the analysis of high-dimensional datasets, such as gene expression profiles, protein folding simulations, and pathway activity models~\cite{huang2023deepmom}. These analyses frequently involve biological replicates with different levels of noise, and MFV is a stable way to determine representative values that are not overly influenced by outliers. For instance, RNA-seq or proteomics data often contain both biological and technical variability~\cite{Evans.etalSelectingBetweensampleRNASeq2018}, and MFV can help extract meaningful trends under such conditions.

In biomedical diagnostics, sensor readings, imaging data, and molecular assays can be affected by noise, sampling inconsistencies, or measurement anomalies. MFV allows these values to be processed more reliably. When combined with bootstrapping--a method that estimates statistical confidence intervals by repeatedly resampling the dataset--the MFV approach provides robust estimates of variability without assuming a specific data distribution~\cite{Idrees.ZhengLowCostAir2020,Bernasconi.etalScopingReviewWearable2022}. This combination is particularly useful when working with small or irregularly sampled datasets, which often occurs in clinical research.

In bioprocess engineering, where biological production systems are optimized and monitored, variability in feedstock, growth rates, and process parameters can complicate the analysis of production yields or reaction efficiencies~\cite{Hutchinson2014,McDonnell22}. The MFV method can help identify representative performance metrics in the presence of such variations, whereas bootstrapping provides reliable uncertainty estimates to support process decisions and quality control.

Bootstrapping is especially useful when the data distribution is unknown or non-Gaussian. Traditional confidence interval (CI) estimation methods~\cite{NichollsConfidenceLimitsError2014} may not perform well under these conditions. Bootstrapping addresses this by generating distributions of the statistic of interest from the resampled datasets, thereby making it highly flexible. When used with MFV, this technique supports statistically sound conclusions even in the presence of extreme values or limited sample sizes.

This study presents a practical approach for estimating confidence intervals by integrating MFV with both traditional and hybrid parametric bootstrapping. The proposed method is designed to be broadly applicable, particularly in cases where datasets are small, variable, or contain outliers. Although the method is demonstrated using a dataset from nuclear physics--specifically, activation cross-section measurements--the statistical issues addressed here are directly relevant to biomolecular sciences. These include challenges encountered in protein-ligand binding assays, enzymatic rate measurements, and molecular biomarker evaluations, such as high-sensitivity C-reactive protein measurements in liver disease studies~\cite{Baek2024}--where population variability, sex-based differences, and limited measurement precision can introduce noise and uncertainty into the analysis.

Historically, before the 1960s, the concept of the most frequent value was known but was rarely used because of computational constraints. The arithmetic mean and Gaussian-based least squares methods were more common even though many real-world datasets did not meet the assumptions required by these methods.

A shift began in the 1970s and 1980s, when researchers such as Steiner, Cserny\'ak
, and Hajagos formalized the MFV method and demonstrated its practical advantages. The key contributions from this period include:
\begin{itemize}
	\item Cserny\'ak
 and Steiner (1980)~\cite{csernyak1980practical}, who introduced a practical way to compute the MFV's scale parameter, known as dihesion.
	\item Steiner (1980)~\cite{steiner1980mfitting}, who demonstrated that MFV offers greater resistance to outliers than traditional least squares fitting.
	\item Cserny\'ak, Hajagos, and Steiner (1981)~\cite{csernyak1981lawoflargenumbers}, who established mathematical foundations for the convergence of MFV estimates.
	\item Hajagos (1982)~\cite{hajagos1982mostfrequentvalue}, who demonstrated that the proposed method minimizes information loss when estimating central values.
\end{itemize}

These developments were compiled in Steiner's
 1988 monograph, \textit{Most Frequent Value Procedures}~\cite{SteinerMostFrequentValue1988}, which helped expand the use of MFV beyond geophysics and into other scientific domains.

This study builds on that foundation by showing how MFV and bootstrapping can be combined to improve data analysis in biomolecular research. Together, they provide a statistically robust framework for analyzing datasets that are small, noisy, or include extreme values--features commonly found in molecular biology, bioassays, and clinical studies. The proposed method enhances confidence in statistical inferences without compromising the integrity of the original data.

\hl{Artificial datasets have been used in past studies to demonstrate the performance of MFV algorithm. For instance, Golovko et al. (2023)}~\cite{golovko2023unveiling} \hl{compared the mode statistic with the MFV statistic using an artificial dataset, and Golovko (2025)}~\cite{golovko2025hpb_IS} \hl{introduced the MFV-hybrid parametric bootstrapping framework with a small four-element dataset that had varied uncertainties. However, in our study, we chose real-world nuclear physics data. We selected this dataset due to its natural variability and tendency to contain outliers, making it an excellent choice for testing the robustness and applicability of the proposed framework. Using real data not only proves the method’s effectiveness but also showcases its potential usefulness across different scientific fields.}

\section{ Methodology }

This study emphasizes the application of robust statistical methods, particularly the most frequent value approach~\cite{Steiner1973}, as an alternative to traditional averaging techniques based on the least squares principle~\cite{Zyla2020}. The MFV method addresses the key limitations inherent in standard methods, such as the arithmetic mean, which assumes that the underlying data follow a Gaussian (normal) distribution. In many real-world scenarios, error distributions significantly deviate from Gaussian assumptions, rendering traditional methods inefficient and prone to inaccuracies.

Traditional methods, such as the arithmetic mean, are optimal only when the data follow a Gaussian distribution. However, non-Gaussian distributions require substantially more data to achieve a similar level of accuracy compared to robust alternatives~\cite{SteinerMostFrequentValue1988}. In addition, least squares techniques are highly sensitive to outliers--data points that deviate significantly from the main data cluster. These outliers can disproportionately influence the resulting estimates, leading to skewed or misleading results.

The MFV approach overcomes these limitations by identifying the densest cluster of data points~\cite{Tolner2024}, effectively reducing the influence of outliers. Unlike traditional averaging methods, the MFV is designed to handle non-Gaussian error distributions~\cite{zhang2024most} and provides more reliable central tendency estimates for datasets with irregular or skewed distributions.

The MFV method offers several notable advantages over traditional averaging techniques, such as the arithmetic mean. A key benefit of an MFV is its robustness to outliers~\cite{golovko2023Ar39}. Unlike traditional methods, which can be heavily influenced by extreme data points, the MFV is based on the concept of minimizing information loss~\cite{zhang2022mfv}, ensuring that the central estimate accurately reflects the majority of the data. This makes it particularly effective in scenarios where datasets contain irregular or extreme values~\cite{Szabo.etalMostFrequentValuebased2018}. In addition, the MFV applies to non-Gaussian distributions, which are frequently encountered in many practical fields. Traditional methods often assume Gaussian error distributions, limiting their effectiveness when this assumption is not met. The MFV, on the other hand, is well-suited for datasets with non-standard distributions~\cite{zhang2018most}.

Another significant advantage of the MFV is its efficiency. By concentrating on the densest cluster of data~\cite{Tolneretal2023}, the MFV provides accurate estimates with fewer data points, thereby reducing the need for extensive sampling. Furthermore, the proposed method improves accuracy by avoiding biases introduced by extreme values, thereby producing results that are closer to the true characteristics of the data. These features make the proposed MFV a significant improvement over traditional least squares techniques, particularly in applications where data variability and non-standard distributions are common~\cite{ZhangMostFrequentValue2017}. The MFV is a robust, efficient, and reliable alternative for statistical analysis, making it an essential tool for modern data analysis.

Although the MFV method and its scale parameter (also referred to as ``dihesion") were originally developed by Steiner and have been discussed in various papers~\cite{GOLOVKO2024143910} and books~\cite{SteinerMostFrequentValue1991a,SteinerOptimumMethodsStatistics1997a}, many of these resources are written for readers already familiar with the approach. As a result, they often skip the step-by-step derivation of the key equations used to calculate the central value and spread. This can make it difficult for researchers new to the MFV method to fully understand how it works or how to apply it effectively.

To address this gap, we have included a detailed explanation of how these equations are derived. The following section describes the process based on the principle of minimizing information loss between the observed dataset and a substituting analytical distribution. Our goal is to make the MFV method more accessible, transparent, and easier to apply in modern statistical analysis, especially when working with small datasets or those containing outliers.

The MFV method, as a robust statistical estimate~\cite{golovko2023unveiling}, is particularly powerful when paired with bootstrapping techniques to estimate confidence intervals (CIs) with high reliability. This combination ensured that the CI was unaffected by the presence of outliers in the dataset. Unlike the mean statistic, which is not robust and can be significantly distorted by extreme values, the MFV minimizes the influence of outliers by focusing on the densest cluster of data. This makes bootstrapping the MFV-based CI a useful approach for datasets with irregular distributions or outliers. By leveraging this robustness, the MFV method provides more reliable and accurate statistical inferences, making it an effective alternative to traditional mean-based methods for estimating CIs.

\subsection{ The most frequent value }

The Cauchy distribution~\cite{CauchyResultatsMoyensDobservations1853}, used in this section is referred to by various names, especially in physics, such as the Lorentz distribution~\cite{LorentzAbsorptionEmissionLines1906}, the Cauchy-Lorentz distribution, the Lorentzian function, or the Breit-Wigner distribution~\cite{BreitWigner1936}. The importance of the Cauchy function lies in its role in most frequent value calculations. It allows for the preservation of information accuracy by substituting an unknown probability distribution with a known one. This practice is based on the principles of information theory and is specifically centered around the idea of \emph{Kullback--Leibler divergence}~\cite{Kullback-Leibler1951}. This divergence measures information loss when a theoretical or analytical distribution approximates the true (but unknown) distribution.

The Cauchy distribution provides several advantages as the substituting distribution in this framework. One significant advantage of the proposed method is its ability to be applied to a broad spectrum of possible true distributions, encompassing those with more pronounced tails than the Gaussian distribution. Consider the scenario where we have two statistically independent random variables, $A$ and $B$ with standard Gaussian distributions. When we calculate the ratio of $A$ to $B$, the resulting distribution of this ratio is actually in the Cauchy form, which is a somewhat unexpected outcome~\cite{WalckHandbookStatisticalDistributions2007}. Unlike the Gaussian distribution, which may lead to infinite divergence for some datasets, the Cauchy distribution ensures finite Kullback--Leibler divergence. This makes it particularly effective for handling nonstandard distributions, which are commonly encountered in real-world data.

Minimizing the Kullback--Leibler divergence with the Cauchy distribution results in defining equations for the MFV (as the location parameter) and dihesion (as the scale parameter). The Cauchy distribution also acts as an ideal weight function in these calculations. It assigns higher weights to data points near the central cluster and progressively downweight the influence of outliers, which enhances the robustness of the MFV estimation. This property ensures that the MFV captures the true central tendency of the data without distortion by extreme values.

Although alternative approaches such as the maximum likelihood principle can also be used to derive MFV and dihesion formulas~\cite{SteinerMostFrequentValue1988}, they rely on the assumption that the true error distribution is known. The Kullback--Leibler divergence method acknowledges the uncertainty inherent in most practical scenarios and focuses on finding the best substituting distribution.

In addition, the Cauchy distribution offers another crucial advantage: its robustness ensures a finite asymptotic variance for the MFV~\cite{hajagos1982mostfrequentvalue,csernyak1982remark}, even in datasets with significant outliers. This makes MFV calculations based on the Cauchy function a highly reliable and practical choice for estimating central tendencies in diverse datasets.

The concept of minimizing information loss addresses a fundamental challenge in data analysis: approximating an unknown true probability distribution with a known analytical distribution. This approach is especially important in fields of science, where the true error distribution is often unknown, and deviations from standard distributions such as the Gaussian distribution, are common. By quantifying and minimizing information loss, scientists can make more informed decisions and extract reliable results from limited or uncertain datasets.

A key tool in this framework is the Kullback--Leibler (KL) divergence (also called relative information entropy~\cite{JaynesInformationTheoryStatistical1957,VerduCauchyDistributionInformation2023} and I-divergence~\cite{Csiszar1975,SteinerMostFrequentValue1988}), a measure of the difference between the actual distribution,  \( f(x) \), and a substituting distribution,  \( g(x) \). Mathematically, Kullback--Leibler divergence is expressed as
\begin{equation}
	D_{\text{KL}}\left(f(x)\Vert g(x)\right) = \int_{-\infty}^\infty f(x) \log \left( \frac{f(x)}{g(x)} \right) \, dx,
	\label{eq:I_divergence_general}
\end{equation}
where \( f(x) \) is the true distribution (unknown), and \( g(x) \) is the substituting distribution, which depends on parameter \( x_0 \), the location parameter. The goal is to minimize \( D_{\text{KL}}\left(f(x)\Vert g(x)\right) \) with respect to \( x_0 \). The Kullback--Leibler divergence shows how different the two functions \( g(x) \) and \( f(x) \) are by measuring the amount of information lost when \( g(x) \) is used to approximate \( f(x) \). While often called statistical ``distance,'' it is not a true distance because it is not symmetric and does not satisfy triangle inequality.

Let \( g(x; x_0, \gamma) \) be the Cauchy distribution~\cite{WalckHandbookStatisticalDistributions2007,VerduCauchyDistributionInformation2023}:
\begin{equation}
	g(x; x_0, \gamma) = \frac{1}{\pi} \cdot \frac{\gamma}{\gamma^2 + (x - x_0)^2},
	\label{eq:cauchy_distribution}
\end{equation}
where \( \gamma \) is the scale parameter (also called half-width at half-maximum), and \( x_0 \) represents the location parameter. The conditions for minimizing the quasi-distance between  \( g(x) \) and  \( f(x) \), as measured by the Kullback--Leibler divergence, are fulfilled if the following equations hold:
\begin{equation}
	\frac{d D_{\text{KL}}\left(f(x)\Vert g(x)\right)}{dx_0} = 0,
	\label{eq:first_derivative_condition}
\end{equation}
and
\begin{equation}
	\frac{d^2 D_{\text{KL}}\left(f(x)\Vert g(x)\right)}{dx_0^2} > 0.
	\label{eq:second_derivative_condition}
\end{equation}
Using the expression for \( D_{\text{KL}}\left(f(x)\Vert g(x)\right) \) from Eq.~(\ref{eq:I_divergence_general}), the condition in Eq.~(\ref{eq:first_derivative_condition}) becomes:
\begin{equation}
	\int_{-\infty}^\infty \frac{\partial g(x; x_0, \gamma)}{\partial x_0} \cdot \frac{f(x)}{g(x; x_0, \gamma)} \, dx = 0,
	\label{eq:first_integral_condition}
\end{equation}
and the second-derivative condition in Eq.~(\ref{eq:second_derivative_condition}) requires:
\begin{equation}
	\int_{-\infty}^\infty \left[ \frac{\partial g(x; x_0, \gamma)}{\partial x_0} \cdot \frac{1}{g(x; x_0, \gamma)} \right]^2 f(x) \, dx
	- \int_{-\infty}^\infty \frac{\partial^2 g(x; x_0, \gamma)}{\partial x_0^2} \cdot \frac{f(x)}{g(x; x_0, \gamma)} \, dx > 0.
	\label{eq:2_integral_condition}
\end{equation}
Equation~(\ref{eq:2_integral_condition}) is automatically fulfilled if the second term vanishes, in other words,
\begin{equation}
	\int_{-\infty}^\infty \frac{\partial^2 g(x; x_0, \gamma)}{\partial x_0^2} \cdot \frac{f(x)}{g(x; x_0, \gamma)} \, dx = 0,
	\label{eq:second_term_zero}
\end{equation}
and the simultaneous satisfaction of Eqs.~(\ref{eq:first_integral_condition}) and (\ref{eq:second_term_zero}) results in values of \( x_0 \) that guarantee the minimum Kullback--Leibler divergence.

For the Cauchy distribution, the partial derivative of \( g(x; x_0, \gamma) \) with respect to \( x_0 \) is given by:
\begin{equation}
	\frac{\partial g(x; x_0, \gamma)}{\partial x_0} = -\frac{2(x - x_0)}{\pi} \cdot \frac{\gamma}{\left[ \gamma^2 + (x - x_0)^2 \right]^2}.
	\label{eq:derivative_cauchy}
\end{equation}
Substituting this expression into the integral condition~(\ref{eq:first_integral_condition}) yields:
\begin{equation}
	\int_{-\infty}^\infty \frac{x - x_0}{\gamma^2 + (x - x_0)^2} f(x) \, dx = 0.
	\label{eq:integral_condition}
\end{equation}
This integral determines the optimal value of \( x_0 \), ensuring that the substituting distribution \( g(x; x_0, \gamma) \) minimizes the Kullback--Leibler divergence relative to the unknown true distribution \( f(x) \). To simplify it, we expand:
\begin{equation}
	\frac{x - x_0}{\gamma^2 + (x - x_0)^2} = \frac{x}{\gamma^2 + (x - x_0)^2} - \frac{x_0}{\gamma^2 + (x - x_0)^2},
	\label{eq:expanded_cauchy_kernel}
\end{equation}
which leads to:
\begin{equation}
	\int_{-\infty}^\infty \left[ \frac{x}{\gamma^2 + (x - x_0)^2} - \frac{x_0}{\gamma^2 + (x - x_0)^2} \right] f(x) \, dx = 0.
	\label{eq:substituted_integral}
\end{equation}
After separating the terms and rearranging them, we isolate \( x_0 \) as
\begin{equation}
	x_0 = \frac{\int_{-\infty}^\infty \frac{x}{\gamma^2 + (x - x_0)^2} f(x) \, dx}{\int_{-\infty}^\infty \frac{1}{\gamma^2 + (x - x_0)^2} f(x) \, dx}.
	\label{eq:T_formula}
\end{equation}
This formula calculates a weighted average of the data, where points close to \( x_0 \) receive larger weights than those farther away. This location parameter, $x_0$ is commonly referred to as $M$ in the MFV methodology.

The next step involves isolating the scale parameter through minimization of the Kullback--Leibler divergence. We use Eq.~(\ref{eq:second_term_zero}) for this purpose. Beginning with the first partial derivative given in Eq.~(\ref{eq:derivative_cauchy}), the second derivative of \( g(x; x_0, \gamma) \) with respect to \( x_0 \) becomes:
\begin{equation}
	\frac{\partial^2 g(x; x_0, \gamma)}{\partial x_0^2} = \frac{2\gamma}{\pi} \cdot \frac{3(x - x_0)^2 - \gamma^2}{\left[ \gamma^2 + (x - x_0)^2 \right]^3}.
	\label{eq:2_partial_der}
\end{equation}
Substituting Eq.~(\ref{eq:2_partial_der}) into Eq.~(\ref{eq:second_term_zero}) leads to
\begin{equation}
	\int_{-\infty}^\infty \frac{3(x - x_0)^2 - \gamma^2}{\left[ \gamma^2 + (x - x_0)^2 \right]^2} f(x) \, dx = 0,
	\label{eq:Q_squared_condition}
\end{equation}
which can be rearranged to isolate \( \gamma^2 \) as:
\begin{equation}
	\gamma^2 = 3 \cdot \frac{ \int_{-\infty}^\infty \frac{(x - x_0)^2}{\left[ \gamma^2 + (x - x_0)^2 \right]^2} f(x) \, dx}{\int_{-\infty}^\infty \frac{1}{\left[ \gamma^2 + (x - x_0)^2 \right]^2} f(x) \, dx}.
	\label{eq:Q_squared_value}
\end{equation}
In the MFV framework, this scale parameter is typically referred to as $\varepsilon$ in place of $\gamma$.

For a finite-sample dataset $\{x_1, x_2, \dots, x_n\}$, the empirical distribution function
\begin{equation}
	f(x) = \frac{1}{n} \sum_{i=1}^{n} \delta(x - x_i)
	\label{eq:empirical_distribution}	
\end{equation}
is used. Substituting the empirical function into Eqs.~(\ref{eq:T_formula}) and (\ref{eq:Q_squared_value}) provides the sample-based MFV, $M_n$, and dihesion, $\varepsilon_n$, through:
\begin{equation}
	M_n = \frac{\sum_{i=1}^{n} \frac{x_i}{\varepsilon_n^2 + (x_i - M_n)^2}}{\sum_{i=1}^{n} \frac{1}{\varepsilon_n^2 + (x_i - M_n)^2}},
	\label{eq:MFV_empirical}
\end{equation}
and
\begin{equation}
	\varepsilon_n^2 = 3 \cdot \frac{\sum_{i=1}^{n} \frac{(x_i - M_n)^2}{\left[ \varepsilon_n^2 + (x_i - M_n)^2 \right]^2}}{\sum_{i=1}^{n} \frac{1}{\left[ \varepsilon_n^2 + (x_i - M_n)^2 \right]^2}},
	\label{eq:dihesion_empirical}
\end{equation}
which define $M_n$ and $\varepsilon_n$ through weighted sums that strongly favor points close to $M_n$, effectively suppressing the influence of outliers.

The equations for $M_n$ and $\varepsilon_n$ each depend on one another's current estimates, so an iterative procedure is required to solve them. Commonly, one starts with:
\begin{align}
	M_n^{(0)} &= \frac{1}{n} \sum_{i=1}^{n} x_i, \\
	\varepsilon_n^{(0)} &= \frac{\sqrt{3}}{2} \cdot (x_{\text{max}} - x_{\text{min}}),
	\label{eq:initial_estimates_updated}
\end{align}
or one can choose the median of the dataset for $M_n^{(0)}$ if the data contain extreme outliers~\cite{HajagosSteiner1992}. The iterative updates are as follows:
\begin{align}
	M_n^{(k+1)} &= \frac{\sum_{i=1}^{n} \frac{x_i}{\left( \varepsilon_n^{(k)} \right)^2 + \left( x_i - M_n^{(k)} \right)^2}}{\sum_{i=1}^{n} \frac{1}{\left( \varepsilon_n^{(k)} \right)^2 + \left( x_i - M_n^{(k)} \right)^2}}, \\
	\left( \varepsilon_n^{(k+1)} \right)^2 &= 3 \cdot \frac{\sum_{i=1}^{n} \frac{(x_i - M_n^{(k)})^2}{\left[ \left( \varepsilon_n^{(k)} \right)^2 + \left( x_i - M_n^{(k)} \right)^2 \right]^2}}{\sum_{i=1}^{n} \frac{1}{\left[ \left( \varepsilon_n^{(k)} \right)^2 + \left( x_i - M_n^{(k)} \right)^2 \right]^2}},
\end{align}
until the changes in both $M_n$ and $\varepsilon_n$ become negligible.

The final values, $M_n$ and $\varepsilon_n$, provide robust estimates of the central tendency (location) and spread (scale) of the dataset. Unlike traditional means and standard deviations, these quantities are far less affected by outliers or heavy-tailed distributions.

It can be helpful to see why Eqs.~(\ref{eq:MFV_empirical}) and (\ref{eq:dihesion_empirical}) look like weighted averages and how they confer robustness. Minimizing Kullback--Leibler divergence with a Cauchy form imposes a higher weight on data points that lie close to the current location estimate, $M_n$. The term $\varepsilon_n^2 + (x_i - M_n)^2$ in the denominator becomes large for points lying far from $M_n$; thus, distant points carry less influence on the estimates. Consequently, outliers do not inflate the location or spread values as strongly as they would in the case of standard arithmetic means or variances.

Intuitively, the most frequent value $M_n$ emerges as the ``peak'' of a Cauchy curve positioned to match the core mass of the data, with minimal impact from points on the periphery. Similarly, the dihesion $\varepsilon_n$ quantifies how widely the main cluster of data is spread around $M_n$, again downlighting extreme values that might otherwise dominate a conventional variance calculation. As a result, these MFV estimates remain stable and representative even when the dataset exhibits strong deviations from Gaussian assumptions or contains significant outliers.

By iteratively updating $M_n$ and $\varepsilon_n$ until convergence, we obtain final values that minimize information loss under Cauchy substitution. This procedure directly addresses the requirement to handle non-Gaussian errors and ensures that the substitution remains valid for a broad range of real-world data distributions. This robustness makes the MFV a valuable alternative to traditional least squares or mean-based methods, offering more reliable central estimates and spread measurements in diverse applications.

It is important to emphasize that the function \( f(x) \) represents the unknown ``true'' distribution of the real-world data. In practice, this distribution can be arbitrary and does not need to follow any standard or symmetric shapes, such as Gaussian. The MFV approach addresses this general case by substituting an unknown distribution with a Cauchy distribution in a way that minimizes Kullback--Leibler divergence. This ensures that the estimated location (and scale) parameters remain robust even when the data are drawn from a heavy-tailed or skewed distribution.

\subsection{ Bootstrapping for Robust Confidence Interval Estimation
}

Bootstrapping is a nonparametric resampling technique used to estimate the variability and confidence intervals of a statistical measure without making strong assumptions about the underlying data distribution~\cite{efron1994introduction,davisonBootstrapMethodsTheir1997}. This approach is particularly useful when working with the MFV because it allows for robust confidence interval estimation while minimizing sensitivity to outliers and small sample sizes.

The procedure generates multiple resampled datasets (termed ``bootstrap samples'') by randomly sampling the original dataset with replacement. For each bootstrap sample, the statistic of interest, such as the MFV is recalculated. This results in an empirical distribution of the statistic from which confidence intervals can be derived using the percentile method~\cite{puth2015variety,mokhtar2023confidence}. The confidence interval (CI) is obtained by identifying the appropriate quantiles of the bootstrap distribution. For instance, a 95.45\% CI is defined as:

\begin{equation}
	\text{CI}_{95.45\%} = \left[ Q_{0.02275}, Q_{0.97725} \right],
\end{equation}
where \( Q_{\alpha} \) represents the \( \alpha \)-th quantile of the bootstrap distribution. Similarly, a 68.27\% CI uses the 15.87th and 84.13th percentiles.

This approach offers several advantages. First, it is highly robust~\cite{data9080095} against outliers, particularly when combined with a robust statistic such as the MFV. By incorporating resampling techniques, bootstrapping accounts for data variability while reducing the influence of extreme values on the confidence interval estimation. Second, bootstrapping is flexible and does not require assumptions about the underlying distribution of data. This makes it a valuable tool in real-world applications where the data structure is complex or unknown.

For small datasets that include measurement uncertainties, the hybrid parametric bootstrap (HPB) method offers a powerful way to estimate confidence intervals~\cite{golovko2024estimation, Golovko2025HPB}. This approach combines two statistical techniques--non-parametric resampling and parametric simulation--to account for both the natural variability in the data and the uncertainties reported with each measurement.

The HPB process begins with a non-parametric bootstrap step. From the original dataset of \(N\) measurements, a new synthetic dataset is created by randomly sampling with replacement. This means that some measurements may appear more than once, while others may not be selected at all. This resampling mimics the uncertainty about which data points best represent the underlying distribution.

Next, each resampled data point is used to generate a simulated value through a parametric step. For each selected point, a new value is randomly drawn from a normal (Gaussian) distribution centered on the original measurement, with the corresponding reported uncertainty used as the standard deviation. If a simulated value is physically impossible--such as a negative cross-section or half-life--it is discarded and redrawn. This ensures all values remain physically meaningful.

After both steps are completed, the most frequent value is calculated for the simulated dataset. This entire two-step process is repeated many times, generating a distribution of MFV estimates. Confidence intervals, such as the 68.27\% range (equivalent to 1-sigma in Gaussian statistics), are then calculated using the percentile method~\cite{puth2015variety,mokhtar2023confidence}.

By combining realistic uncertainty modeling with repeated resampling, the HPB method provides a statistically robust and physically meaningful way to quantify uncertainty in central estimates, even when working with small or irregular datasets.

The integration of MFV with bootstrapping provides a powerful framework to analyze small datasets, which are costly to obtain, or are prone to outliers. This is especially beneficial in fields such as environmental monitoring and radiation measurement, where robust statistical methods are essential for accurate analysis. By iteratively resampling and recalculating the MFV, the derived confidence intervals become more reliable and representative of the underlying data. This methodology ensures minimal information loss, making it a practical and effective approach for applications requiring high-precision statistical inference.

\section{ Description of the Methods and Results
}

Nuclear physics cross-sectional datasets provide compelling cases for applying robust statistical methods such as the MFV approach and bootstrapping. These datasets often contain inconsistencies, outliers, and non-Gaussian distributions stemming from experimental challenges, including fluctuations in neutron energy, difficulties with sample preparation, and uncertainties in detection methods. As a result, different laboratories may report varying cross-sectional values for the same nuclear reaction due to differences in equipment calibration, experimental procedures, and data processing methodologies. This variability exposes the limitations of conventional statistical tools, especially those that rely on the arithmetic mean, which can be significantly skewed by outliers.

The MFV method offers a more resilient measure of central tendency by identifying the most densely populated data region. When combined with bootstrapping to estimate confidence intervals, this approach yields robust and interpretable results, particularly for datasets with limited sample sizes. The use of advanced statistical techniques in nuclear physics not only improves the reliability of data analysis and enhances the precision of outcomes in critical applications, such as nuclear reactor design, isotope production, and radiation shielding.

As part of this work, a benchmark dataset based on neutron lifetime measurements~\cite{zhang2022mfv} was used to further validate the MFV method. These measurements provide an independent check that the MFV approach yields consistent results. In fact, the same MFV value for neutron lifetime was obtained as reported in the original study~\cite{zhang2022mfv}, confirming the method's accuracy and reproducibility.

The dataset of \(^{109}\text{Ag}(\text{n},2\text{n})^{108m}\text{Ag}\) cross-sections in Table~\ref{tab:sigma_table} serves as a practical example for applying the MFV method and bootstrapping. The experimental setup introduces significant uncertainties and inconsistencies, reflecting its complexity. The MFV method mitigates the influence of outliers, and bootstrapping provides insight into the variability and confidence intervals of the data.

\begin{table}[t]
	\centering
	\caption{Summary of fast-neutron \(E_n = 14.7 \pm 0.2~\mathrm{MeV}\) activation cross-section ($\sigma$) of the $^{109}\text{Ag}(\text{n},2\text{n})^{108\text{m}}\text{Ag}$ reaction.}
	\label{tab:sigma_table}
	\begin{tabular}{cccccc}
		\toprule
		\(E_n~\mathrm{(MeV)}\) & $\sigma$ (mb) & Uncertainty (mb) & References & Year & Comments \\
		\midrule
		14.70 & 670 & 266 & \cite{H._Vonach_et_al_1969} & 1969 & measured \\
		14.50 & 220 &  12 & \cite{Csikai_et_al_1991} & 1991 & measured \\
		14.50 & 263 &  20 & \cite{Csikai_et_al_1991} & 1991 & measured \\
		14.83 & 767 &  24 & \cite{Lu_et_al_1991} & 1991 & evaluation \\
		14.60 & 232 &   8 & \cite{Wang_et_al_1992} & 1992 & measured \\
		14.80 & 236 &   7 & \cite{Wang_et_al_1992} & 1992 & measured \\
		14.70 & 628 &  42 & \cite{Meadows_et_al_1996} & 1996 & measured \\
		14.70 & 682 &  49 & \cite{Meadows_et_al_1996} & 1996 & measured \\
		14.50 & 697 &  60 & \cite{Qaim_et_al_1996} & 1996 & measured \\
		14.50 & 621 &  29 & \cite{Ikeda_et_al_1996} & 1996 & evaluation \\
		14.80 & 648 &  31 & \cite{Ikeda_et_al_1996} & 1996 & evaluation \\
		14.80 & 721 &  20 & \cite{Csikai1996} & 1996 & evaluation \\
		14.50 & 695 &  40 & \cite{pashchenko1993activation} & 1997 & evaluation \\
		14.90 & 709 &  41 & \cite{pashchenko1993activation} & 1997 & evaluation \\
		14.50 & 643 &  30 & \cite{pashchenko1993activation} & 1997 & evaluation \\
		14.80 & 671 &  31 & \cite{pashchenko1993activation} & 1997 & evaluation \\
		14.70 & 651 &  44 & \cite{pashchenko1993activation} & 1997 & evaluation \\
		14.70 & 706 &  51 & \cite{pashchenko1993activation} & 1997 & evaluation \\
		14.50 & 677 &  82 & \cite{pashchenko1993activation} & 1997 & evaluation \\
		14.50 & 716 &  44 & \cite{pashchenko1993activation} & 1997 & evaluation \\
		14.77 & 784 &  25 & \cite{pashchenko1993activation} & 1997 & evaluation \\
		14.83 & 795 &  25 & \cite{pashchenko1993activation} & 1997 & evaluation \\
		14.60 & 790 &  27 & \cite{pashchenko1993activation} & 1997 & evaluation \\
		14.80 & 805 &  24 & \cite{pashchenko1993activation} & 1997 & evaluation \\
		14.80 & 800 &  55 & \cite{Luo_et_al_2009} & 2009 & measured \\
		14.50 & 727 &  41 & \cite{Filatenkov_2016} & 2016 & measured \\
		14.80 & 747 &  42 & \cite{Filatenkov_2016} & 2016 & measured \\
		14.80 & 650 &  43 & \cite{Y._Song_et_al_2024} & 2024 & measured \\
		14.80 & 602 &  40 & \cite{Y._Song_et_al_2024} & 2024 & measured \\
		14.80 & 601 &  37 & \cite{Y._Song_et_al_2024} & 2024 & measured \\
		14.80 & 616 &  23 & \cite{Y._Song_et_al_2024} & 2024 & measured \\
		\bottomrule
	\end{tabular}
\end{table}

Measuring the \(^{109}\text{Ag}(\text{n},2\text{n})^{108m}\text{Ag}\) cross-section remains challenging due to experimental difficulties, technical constraints, and limited resources. Generating and controlling fast-neutrons is particularly problematic. Neutron sources, whether from reactors or generators, can be difficult to monitor precisely in terms of flux and energy. Often, neutron energies are inferred from cross-section ratios of other reactions, introducing additional uncertainty.

Because the cross-section is highly dependent on neutron energy, we restricted our analysis to measurements within the 14.7~\(\pm\)~0.2~MeV range, specifically between 14.5 and 14.9~MeV (see Table~\ref{tab:sigma_table}). This filtering ensures uniformity and reduces variability under different experimental energy conditions.

Sample preparation also affects the measurement quality. High-purity silver samples with well-defined geometry minimize bias~\cite{Y._Song_et_al_2024}. Cadmium shielding is typically employed to suppress low-energy neutron capture on \(^{107}\text{Ag}\), although some residual contributions remain~\cite{Y._Song_et_al_2024}. Gamma-ray spectrometry adds further uncertainty due to factors such as detector efficiency, counting statistics, and self-absorption corrections.

The half-life of \(^{108\text{m}}\text{Ag}\) has undergone significant revision over the past decades, which has also contributed to variations in the reported cross-section values for reactions producing this isomer. Early theoretical studies suggested a minimum half-life of five years~\cite{Meinke1960}, which was subsequently refined by a series of experimental investigations: \(127 \pm 7\) years~\cite{Habbottle1970}, \(418 \pm 15\) years~\cite{Schotzig.et.al.1992}, \(437.7 \pm 7.7\) years~\cite{SCHRADER2004317}, and most recently \(448 \pm 27\) years~\cite{SHUGART2018}. 

A landmark early experimental determination was reported by Vonach et al. in 1969, who estimated the half-life to be \(310 \pm 132\) years~\cite{H._Vonach_et_al_1969}. Their approach relied on evaluating the absolute activity of a sample containing a known quantity of \(^{108\text{m}}\text{Ag}\), and estimating the half-life using the cross-section of the \(^{109}\text{Ag}(\text{n},2\text{n})^{108\text{m}}\text{Ag}\) reaction, for which they adopted a value of \(670 \pm 266~\text{mb}\). Although this measurement carried a large uncertainty, it was based on a methodology distinct from later decay-spectroscopy-based measurements. As such, it provides valuable complementary information rooted in different experimental systematics. This makes the 1969 result a critical reference point for validating updated half-life determinations through modern statistical reanalysis.

\begin{figure}[t]
	\centering
	\includegraphics[width=0.99\textwidth]{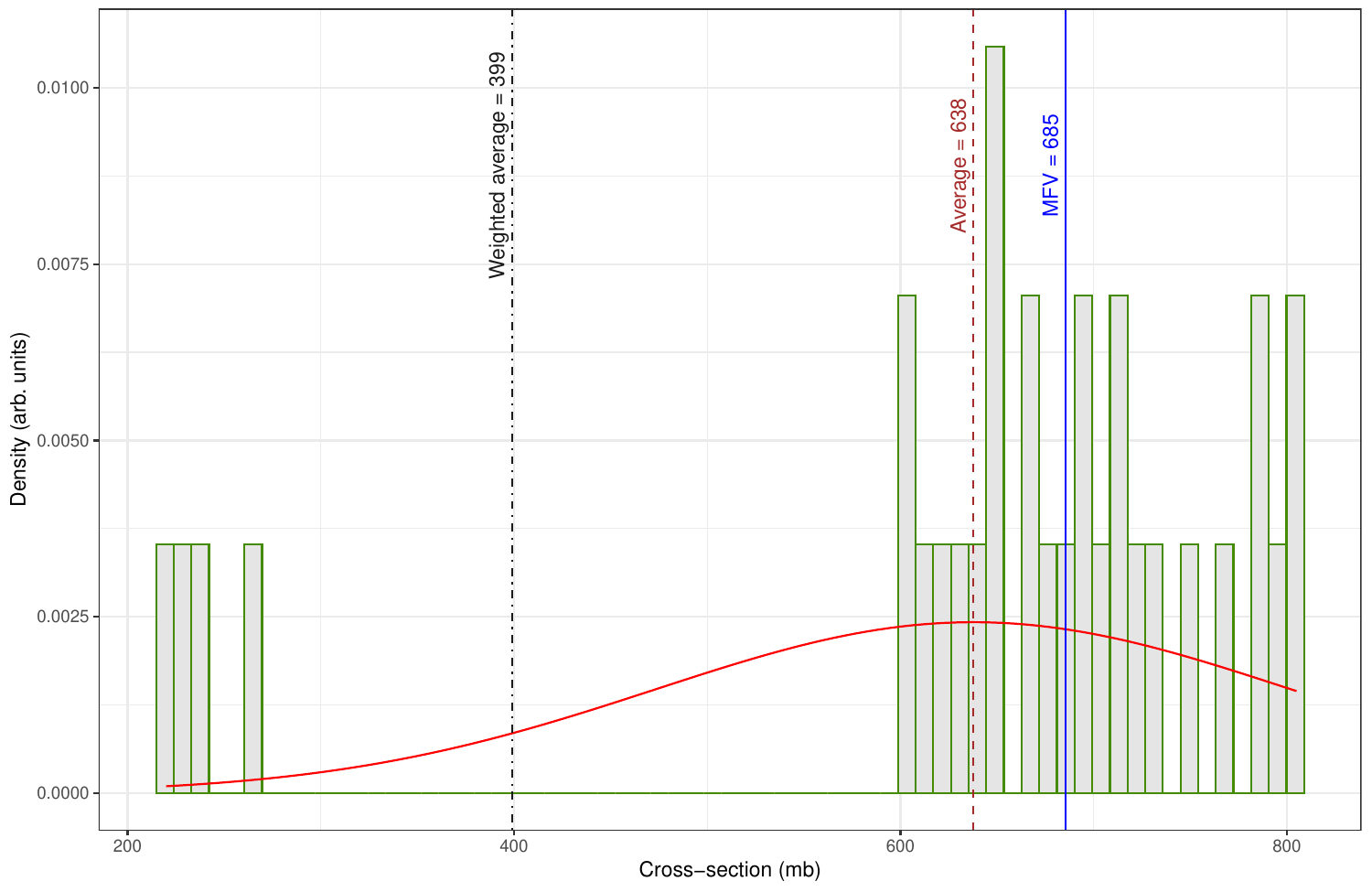}
	\caption{A histogram of fast-neutron (14.7$\pm$0.2~MeV) activation cross-sections (see Table~\ref{tab:sigma_table}) of the \(^{109}\text{Ag}(\text{n},2\text{n})^{108m}\text{Ag}\), showcasing the weighted average (399 mb), arithmetic mean (638 mb), and MFV (685 mb) as measures of central tendency. In addition, the smooth red curve shows the Gaussian fit of the data.}
	\label{fig:histogram}
\end{figure}

In 2024, Y.~Song et al.~\cite{Y._Song_et_al_2024} re-evaluated several fast-neutron activation cross-sections using an updated half-life for \(^{108m}\text{Ag}\) of 438 years, as shown in Table~\ref{tab:Ag108m_14_7MeV}. Their adjustments significantly shifted the data distribution toward higher values. The original low-value cluster between 200--300~mb is absent in the re-evaluated set; instead, it clustered tightly between 600--800~mb, as shown in Figure~\ref{fig:histogram}.

\begin{table}[b]
	\centering
	\caption{Re-evaluation of the fast-neutron cross-section \((\sigma_2)\) for \(^{108\text{m}}\text{Ag}\) at neutron energy \(E_n = 14.7 \pm 0.2~\mathrm{MeV}\), as conducted by Y. Song et al.~\cite{Y._Song_et_al_2024} based on previous measurements \((\sigma_1)\).}
	\label{tab:Ag108m_14_7MeV}
	\begin{tabular}{cccc}
		\toprule
		$E_n$ (MeV) & $\sigma_1$ (mb) & $\sigma_2$ (mb) & Reference \\
		\midrule
		14.70 & $628 \pm 42$ & $658 \pm 44$ & \cite{Meadows_et_al_1996} \\
		14.70 & $682 \pm 49$ & $715 \pm 41$ & \cite{Meadows_et_al_1996} \\
		14.50 & $697 \pm 60$ & $705 \pm 61$ & \cite{Qaim_et_al_1996} \\
		14.50 & $621 \pm 29$ & $651 \pm 30$ & \cite{Ikeda_et_al_1996} \\
		14.80 & $648 \pm 31$ & $679 \pm 32$ & \cite{Ikeda_et_al_1996} \\
		14.80 & $721 \pm 18$ & $755 \pm 21$ & \cite{Csikai1996} \\
		14.60 & $232 \pm 8$  & $800 \pm 28$ & \cite{Wang_et_al_1992} \\
		14.80 & $236 \pm 7$  & $814 \pm 24$ & \cite{Wang_et_al_1992} \\
		14.50 & $220 \pm 12$ & $759 \pm 41$ & \cite{Csikai_et_al_1991} \\
		14.50 & $263 \pm 20$ & $907 \pm 69$ & \cite{Csikai_et_al_1991} \\
		14.83 & $767 \pm 24$ & $804 \pm 25$ & \cite{Lu_et_al_1991} \\
		\bottomrule
	\end{tabular}
\end{table}

\begin{figure}[t]
	\centering
	\includegraphics[width=0.99\textwidth]{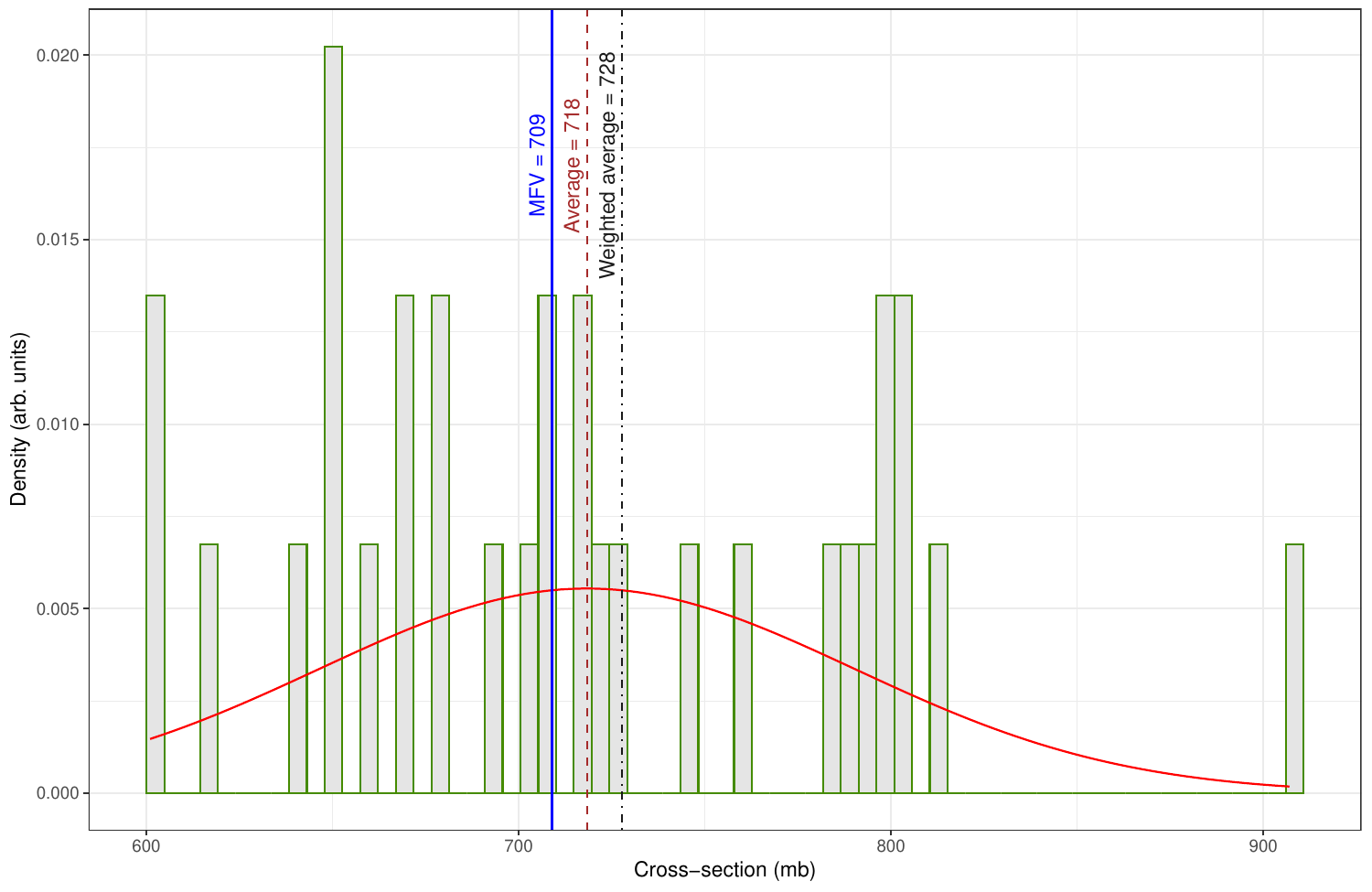}
	\caption{A histogram of fast-neutron (14.7$\pm$0.2~MeV) activation cross-sections (re-evaluated) for the \(^{109}\text{Ag}(\text{n},2\text{n})^{108m}\text{Ag}\) reaction displays the weighted average (728 mb), arithmetic mean (718 mb), and MFV (709 mb) as indicators of central tendency. The smooth red curve shows a Gaussian fit to the data.}
	\label{fig:histogram1}
\end{figure}

Figure~\ref{fig:histogram1}, which presents the re-evaluated fast-neutron activation cross-section data for the \(^{109}\text{Ag}(\text{n},2\text{n})^{108\text{m}}\text{Ag}\) reaction at \(E_n = 14.7 \pm 0.2~\mathrm{MeV}\), showing a marked improvement in data consistency compared to the original dataset displayed in Figure~\ref{fig:histogram}.

In Figure~\ref{fig:histogram}, the distribution of cross-section values is notably irregular and skewed, with a broad spread of measurements ranging from low to high values. This dataset includes several low-valued outliers that significantly influence the statistical estimates. As a result, the weighted average is much lower (\(399~\mathrm{mb}\)) than both the arithmetic mean (\(638~\mathrm{mb}\)) and the MFV (\(685~\mathrm{mb}\)). The mismatch between these indicators of central tendency highlights the presence of inconsistencies and the potential impact of outliers on the overall estimation.

In contrast, Figure~\ref{fig:histogram1} shows a more symmetrical and concentrated distribution, with most cross-section values clustered between approximately \(640\) and \(780~\mathrm{mb}\). The re-evaluated dataset appears to reduce or eliminate the influence of earlier outliers. Consequently, the statistical measures are in close agreement: the weighted average is \(728~\mathrm{mb}\), the arithmetic mean is \(718~\mathrm{mb}\), and the MFV is \(709~\mathrm{mb}\). The improved alignment among these values indicates that the dataset is more internally consistent and statistically robust.

A comparison of central tendency measures between the original and re-evaluated datasets revealed substantial differences, particularly in how each estimator responded to outliers. The arithmetic mean increased from \(638~\mathrm{mb}\) in Figure~\ref{fig:histogram} to \(718~\mathrm{mb}\) in Figure~\ref{fig:histogram1}, representing a percentage difference of approximately 12.5\%. In contrast, the weighted mean underwent a dramatic increase from \(399~\mathrm{mb}\) to \(728~\mathrm{mb}\), corresponding to an 82.5\% difference. This large change indicates that the original dataset contained low-valued outliers that significantly influenced the weighted average, and their impact was effectively mitigated in the updated dataset. The MFV showed only a small change, rising from \(685~\mathrm{mb}\) to \(709~\mathrm{mb}\), with a percent difference of about 3.5\%. The relatively small change observed in the MFV between the two datasets suggests that this estimator is more forgiving with respect to data quality, exhibiting robustness against the presence of outliers or inconsistencies that significantly affect other estimators, such as the mean or weighted average.

This comparison highlights the value of re-evaluating nuclear data using updated nuclear parameters. This study also demonstrates the importance of selecting appropriate statistical estimators--especially MFV--for achieving reliable analysis. By correcting problematic measurements, the updated dataset in Figure~\ref{fig:histogram1} allows for a more accurate estimation of central tendency and supports stronger confidence in subsequent statistical analysis.

Furthermore, we aim to evaluate the concept of calculating confidence intervals using a robust MFV estimator within the HPB framework.The HPB method explicitly incorporates the uncertainty associated with each individual data point. Our goal was to evaluate how sensitive the estimated confidence intervals to the presence of outliers in the dataset. This comparison helps assess the robustness of different estimation approaches under different data quality conditions.

To do this, we estimated the confidence intervals using the HPB method based on the data in Table~\ref{tab:sigma_table} (see Figure~\ref{fig:histogram}). We then repeated the analysis using the re-evaluated data from Y.~Song et al.~\cite{Y._Song_et_al_2024}, shown in Table~\ref{tab:Ag108m_14_7MeV}. We matched rows between the two tables using neutron energy, original cross-sectional values, and associated uncertainties. The matching entries were then updated with the re-evaluated values. This ensures that the final dataset shown in Figure~\ref{fig:histogram1} incorporates the most current and accurate nuclear information, providing a solid foundation for robust confidence interval estimation.

The HPB method, originally proposed in~\cite{golovko2024estimation} and fully described in~\cite{golovko2025hpb_IS}, has been successfully applied in nuclear safety evaluations and environmental monitoring~\cite{Golovko2025HPB}, where data quality varies and outliers are common. In this case, its application is particularly appropriate given the limited dataset and the presence of variable uncertainty levels.

\section{ Discussion }

\begin{figure}[t]
	\centering
	\includegraphics[width=0.99\textwidth]{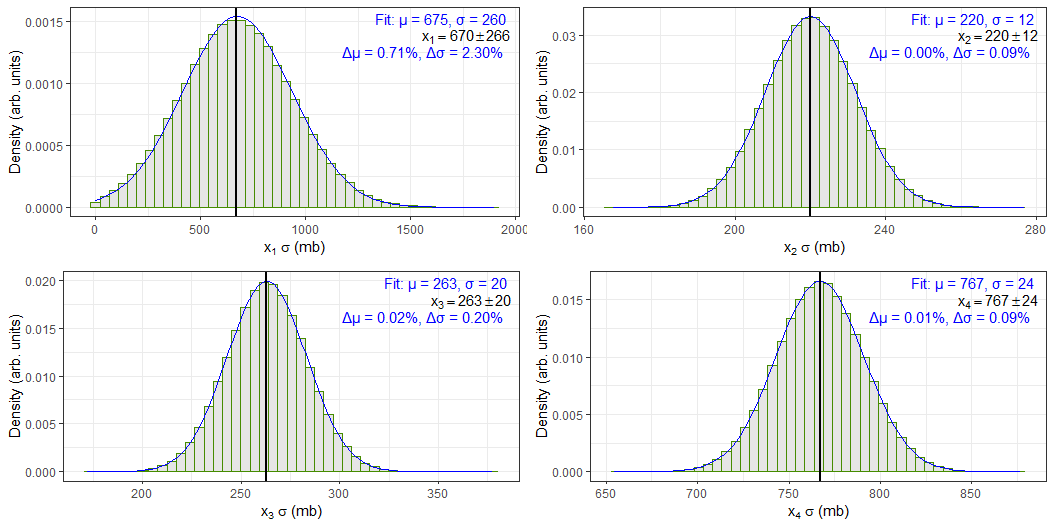}
	\caption{Histograms of randomized bootstrap sample values for four selected cross-section measurements ($x_1$, $x_2$, $x_3$, $x_4$). Each histogram is fitted with a Gaussian function, and the original values with uncertainties from Table~\ref{tab:sigma_table} are marked with vertical black lines. Fit results and absolute percent differences are indicated.}
	\label{fig:bootstrap_histograms}
\end{figure}

To validate the correctness and consistency of the randomized bootstrap sample values used in the MFV-HPB statistical framework, we analyzed a subset of four cross-section entries from Table~\ref{tab:sigma_table}, specifically $x_1$, $x_2$, $x_3$, and $x_4$. These values were randomly sampled using a Gaussian distribution based on their reported central values and uncertainties. Figure~\ref{fig:bootstrap_histograms} presents the histograms of these samples along with fitted Gaussian curves.

To evaluate how closely the randomized bootstrap samples match the original measurements, each histogram was fitted with a Gaussian distribution. The resulting mean ($\mu$) and standard deviation ($\sigma$) from the fit were then compared to the corresponding reference values from Table~\ref{tab:sigma_table}. For each case, the absolute percent difference was calculated to quantify the deviation between the fitted and tabulated values:

\[
\Delta \mu = \left|\frac{\mu_{\text{fit}} - \mu_{\text{ref}}}{\mu_{\text{ref}}}\right| \times 100\%, \quad
\Delta \sigma = \left|\frac{\sigma_{\text{fit}} - \sigma_{\text{ref}}}{\sigma_{\text{ref}}}\right| \times 100\%
\]

Here, $\mu_{\text{fit}}$ and $\sigma_{\text{fit}}$ are the parameters obtained from the fit, while $\mu_{\text{ref}}$ and $\sigma_{\text{ref}}$ are the original values from the table. These percent differences are shown on each panel of Figure~\ref{fig:bootstrap_histograms} and provide a simple measure of consistency between the fit results and the expected values. All differences were found to be small, indicating that the Gaussian randomization process preserved the statistical characteristics of the original data.

It is worth noting that for the $x_1$ dataset, the reported uncertainty in Table~\ref{tab:sigma_table} is comparable in magnitude to the cross-section value itself. This presents a potential challenge when generating randomized bootstrap values using a Gaussian distribution, as the sampling process could yield negative values. Since negative cross-sections have no physical meaning, any such values encountered during the bootstrap process were discarded, and a new random sample was drawn in their place to preserve physical consistency. This correction mechanism can slightly distort the expected shape of the resulting distribution. Consequently, the fitted Gaussian parameters for $x_1$ show a modest deviation from the reference value. However, this discrepancy remains much smaller than the associated uncertainty, and as such, does not compromise the reliability or conclusions drawn from the MFV-HPB analysis.

Overall, the excellent agreement between the Gaussian fits and the original cross-section values affirms that the randomized samples preserve the statistical characteristics of the source data. This outcome confirms the reliability of the bootstrapped datasets as valid inputs for robust estimation using the MFV-HPB framework.

\begin{figure}[t]
	\centering
	\includegraphics[width=0.99\textwidth]{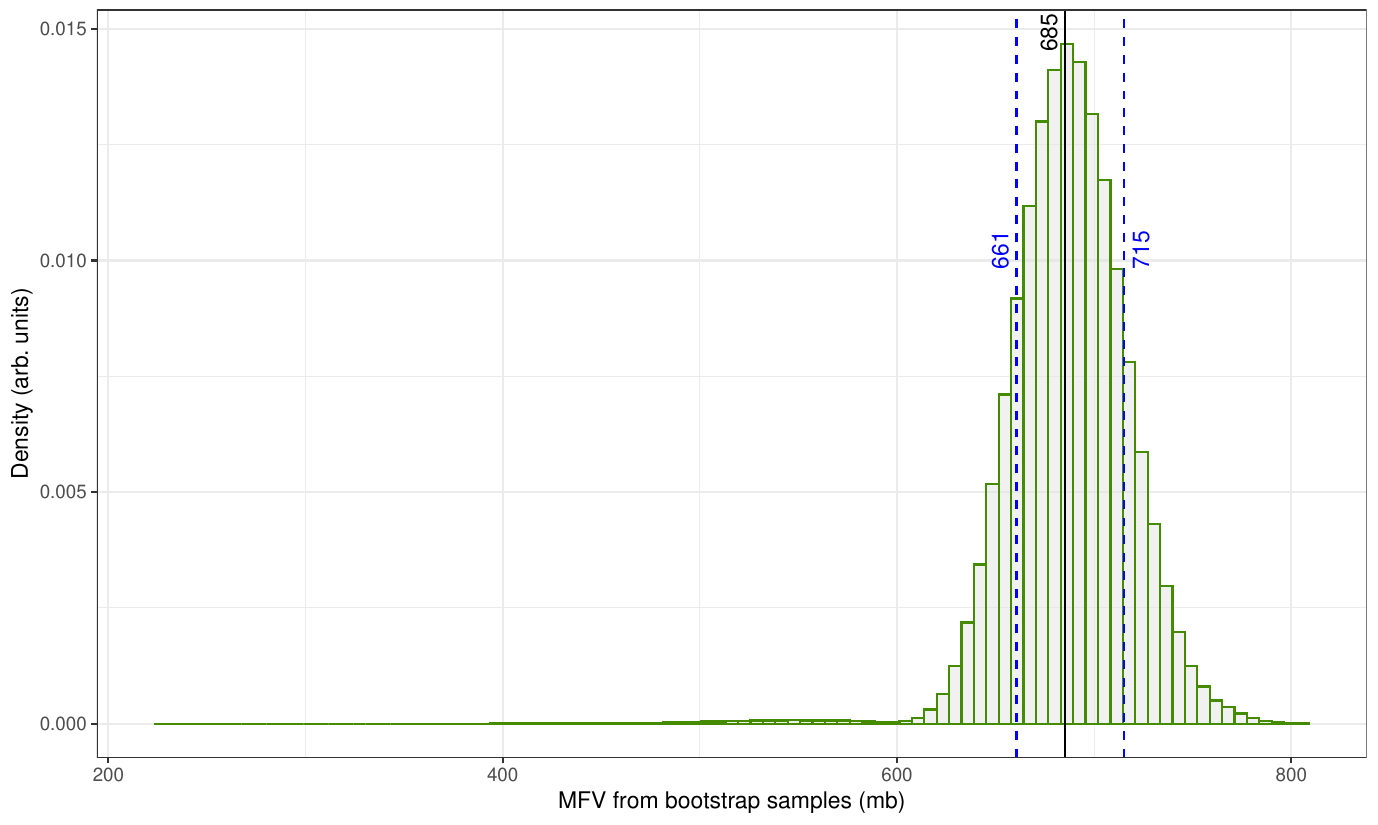}
	\caption{A histogram of the MFV for fast-neutron (14.7$\pm$0.2~MeV) activation cross-sections for the \(^{109}\text{Ag}(\text{n},2\text{n})^{108m}\text{Ag}\) reaction for data from Table~\ref{tab:sigma_table} (original). Hybrid parametric bootstrapping for 68.3\% confidence interval and the MFV are also shown.}
	\label{fig:mfv_bootstrap_hist}
\end{figure}

Figure~\ref{fig:mfv_bootstrap_hist} illustrates the distribution of MFV estimates for the \(^{109}\text{Ag}(\text{n},2\text{n})^{108\text{m}}\text{Ag}\) reaction using the original fast-neutron cross-section dataset from Table~\ref{tab:sigma_table}. These estimates were generated using HPB, a method that incorporates both measurement uncertainties and sampling variability~\cite{Golovko2025HPB}.

The x-axis shows the MFV values obtained from individual bootstrap samples, expressed in millibarns (mb), while the y-axis represents the normalized density. The distribution is centered around the original dataset's MFV of \(685~\mathrm{mb}\), as indicated by a solid vertical line. The 68.27\% confidence interval (marked by dashed blue lines) ranges from approximately \(661~\mathrm{mb}\) to \(715~\mathrm{mb}\). Although not shown in the figure, a broader 2-sigma interval was also calculated to extend from about \(633~\mathrm{mb}\) to \(745~\mathrm{mb}\). The intervals were determined using the percentile method.

To ensure statistical reliability, 500{,}000 bootstrap replicates were generated. This large sample size helps reduce the influence of random fluctuations and improves the precision of the resulting confidence intervals. Additionally, only physically meaningful, positive-valued cross-sectional data were used in the analysis, especially in light of the large uncertainties associated with some older measurements, such as the 1969 value of \(670 \pm 266~\mathrm{mb}\)~\cite{H._Vonach_et_al_1969}.

The resulting histogram is narrow and symmetric with minimal skew. This confirms that the MFV is a stable and reliable central estimate even when applied to a dataset that includes variable uncertainties and possible outliers. The narrow confidence interval further demonstrates that the combined MFV and HPB approach~\cite{golovko2024estimation} provides a robust framework for summarizing central tendency and uncertainty in nuclear cross-sectional data.

\begin{figure}[t]
	\centering
	\includegraphics[width=0.99\textwidth]{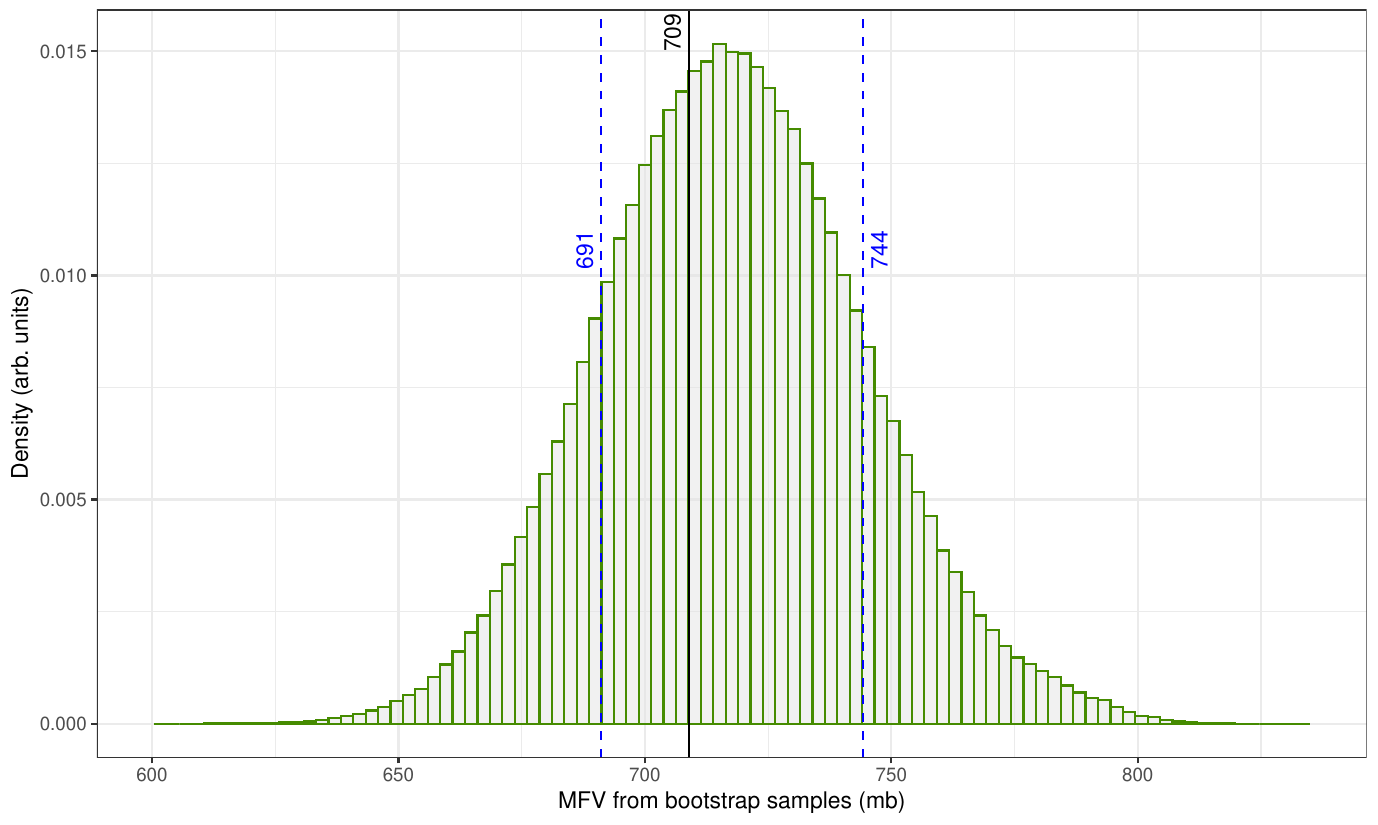}
	\caption{A histogram of the MFV for fast-neutron (14.7$\pm$0.2~MeV) activation cross-sections for the \(^{109}\text{Ag}(\text{n},2\text{n})^{108m}\text{Ag}\) reaction for data from Table~\ref{tab:Ag108m_14_7MeV} (re-evaluated). Hybrid parametric bootstrapping for 68.27\% confidence interval and the MFV are also shown.}
	\label{fig:mfv_bootstrap_revised}
\end{figure}

Figure~\ref{fig:mfv_bootstrap_revised} presents a similar analysis applied to the re-evaluated dataset with the updated cross-sectional values from Table~\ref{tab:Ag108m_14_7MeV}. This dataset incorporates updated nuclear information, including a revised half-life for \(^{108\text{m}}\text{Ag}\), and yields an MFV centered at approximately \(709~\mathrm{mb}\). The histogram again displays the distribution of MFV values derived from HPB resampling.

The shape of the distribution in Figure~\ref{fig:mfv_bootstrap_revised} is slightly right-skewed, with the 68.27\% confidence interval spanning from \(691~\mathrm{mb}\) to \(744~\mathrm{mb}\). The corresponding 2-sigma interval (not shown) ranges from \(666~\mathrm{mb}\) to \(774~\mathrm{mb}\). Compared with the original data in Figure~\ref{fig:mfv_bootstrap_hist}, the revised dataset produced slightly higher MFV estimates while maintaining a similar spread. This shift reflects a higher concentration of consistent values in the upper cross-section range, as seen in the re-evaluated measurements (see Figure~\ref{fig:histogram1}).

The stability of the MFV across both the original and re-evaluated datasets underscores the strength of the proposed method. While conventional metrics like the mean or weighted average, are easily distorted by extreme values or skewed uncertainties, the MFV remains centered within the most densely populated region of the data. By combining MFV with HPB, we can generate confidence intervals that are not only statistically rigorous but also not sensitive to the quality and distribution of the original measurements.

These findings highlight the advantages of MFV-based analysis of nuclear data, particularly in the presence of heterogeneous uncertainty and limited experimental repetition. The proposed method reduces the bias introduced by outliers and leverages resampling to quantify uncertainty without assuming a specific underlying distribution. Although computationally intensive, the MFV and HPB approaches are well-suited for nuclear datasets where precision matters and data collection is costly.

Together, Figures~\ref{fig:mfv_bootstrap_hist} and~\ref{fig:mfv_bootstrap_revised} demonstrate that combining the MFV estimator with hybrid bootstrapping results in a consistent and interpretable analysis framework. This methodology offers a promising direction for future nuclear data evaluation efforts, especially when dealing with incomplete, conflicting, or high-uncertainty datasets.

For datasets containing more than 10 elements, the use of nonparametric bootstrapping is generally considered appropriate for estimating statistical confidence intervals~\cite{golovko2025smart}. Singh et al.~\cite{singhComputation95Upper2006a} recommend this guideline. In this study, we used a dataset of 31 values, which would normally allow for such nonparametric bootstrap. However, we chose the hybrid parametric bootstrap method instead. Unlike traditional resampling, HPB accounts for the measurement uncertainty of each data point and incorporates it directly into the resampling process. As a result, it produces confidence intervals that are more representative of real-world conditions. In applications such as nuclear cross-section calculations, it is also essential to ensure that generated values remain physically valid--for example, cross-sections must be strictly positive. The HPB method allows such constraints to be imposed as part of its core procedure.

To evaluate the robustness of the MFV-HPB method when applied to small datasets, we analyzed five published half-life values for \(^{108\text{m}}\text{Ag}\), each reporting a central estimate and associated uncertainty. These values span nearly five decades of research, ranging from Vonach et al.'s 1969 result to the latest measurements in 2018. The earliest estimate, \(310 \pm 132\) years~\cite{H._Vonach_et_al_1969}, relied on neutron activation and cross-section-based modeling, whereas subsequent values were derived from decay spectroscopy with progressively improved precision.

Using all five values, the MFV analysis produced a half-life of \(433.5~\mathrm{years}\), with a 68.27\% confidence interval (1-sigma) of \([299.2, 440.9]~\mathrm{years}\) and a 95.45\% interval (2-sigma) of \([124.4, 456.7]~\mathrm{years}\). These broad intervals reflect both the limited sample size and the large uncertainty associated with the 1969 measurement.

To test whether the earlier, less precise estimate significantly influenced the result, we repeated the MFV-HPB analysis using only the three most recent and precise values from 1992, 2004, and 2018. This yielded an MFV of \(439.1~\mathrm{years}\), with narrower confidence intervals: \([420.3, 445.3]~\mathrm{years}\) for 1-sigma and \([404.1, 466.5]~\mathrm{years}\) for 2-sigma.

The difference between the two MFV results was modest--less than \(6~\mathrm{years}\)--and well within the respective uncertainty ranges. Notably, the second analysis excluded not only the earliest measurement by Vonach et al.~\cite{H._Vonach_et_al_1969} (\(310 \pm 132~\mathrm{years}\)), but also the 1970 value of \(127 \pm 7~\mathrm{years}\)~\cite{Habbottle1970}, which is significantly lower than later estimates and based on earlier-generation measurement techniques. Despite removing these two lower and more uncertain values, the resulting MFV changed only slightly.

\begin{figure}[t]
	\centering
	\includegraphics[width=0.99\textwidth]{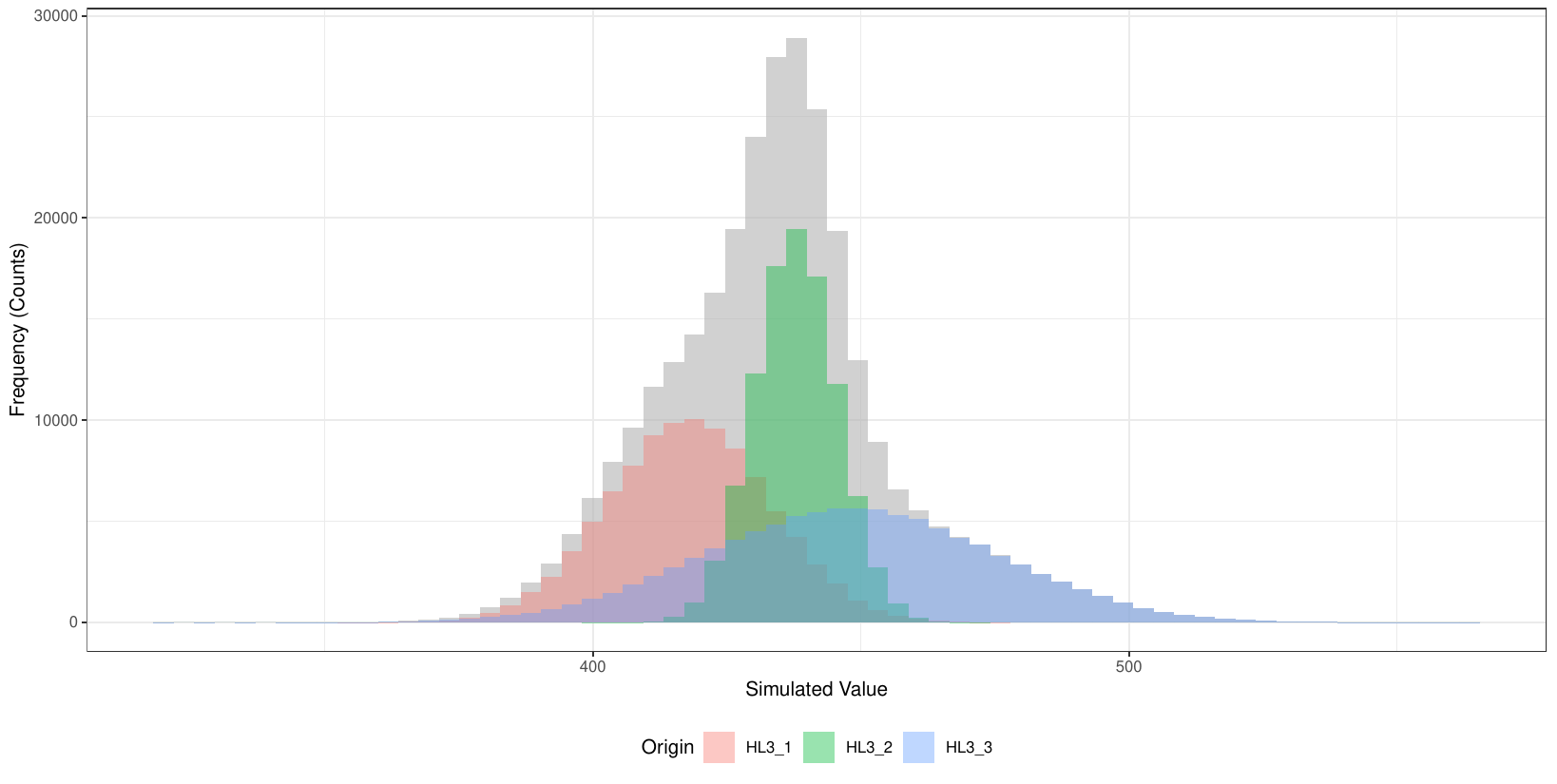}
	\caption{
		Histogram showing the distribution of hybrid parametric bootstrap simulated half-life values for the three most recent measurements of the \(^{108\text{m}}\text{Ag}\) isotope. 
		Each color represents one bootstrapped half-life dataset: \(418 \pm 15\)~years (HL3\_1), \(437.7 \pm 7.7\)~years (HL3\_2), and \(448 \pm 27\)~years (HL3\_3).
		The grey histogram in the background shows the combined distribution of all simulated values.
	}
	\label{fig:bootstrap_histogram}
\end{figure}

\begin{table}[h!]
	\centering
	\caption{
		Comparison between the simulated Gaussian fit results and original bootstrapped half-life values for the three datasets used in the bootstrap analysis. 
		Small absolute percent differences indicate excellent agreement between the simulation and original data.
	}
	\label{tab:bootstrap_fit_summary}
	\begin{tabular}{lcccccc}
		\toprule
Origin & \(N\) & \(\mu\) (sim.) & \(\mu\) (orig.) & \(\sigma\) (sim.) & \(\sigma\) (orig.) & \% Difference (\%) \\
\midrule
HL3\_1 & 99,693 & 417.94 & 418.0 & 14.99 & 15.0 & \(\Delta \mu = 0.01\), \(\Delta \sigma = 0.10\) \\
HL3\_2 & 99,572 & 437.75 & 437.7 & 7.69 & 7.7 & \(\Delta \mu = 0.01\), \(\Delta \sigma = 0.19\) \\
HL3\_3 & 100,735 & 448.05 & 448.0 & 27.11 & 27.0 & \(\Delta \mu = 0.01\), \(\Delta \sigma = 0.41\) \\
\bottomrule
	\end{tabular}
\end{table}

The bootstrap values for the hybrid parametric bootstrap analysis, focusing on the three latest half-life measurements of \(^{108\text{m}}\text{Ag}\), are shown in Figure~\ref{fig:bootstrap_histogram}. A summary of these values is also provided in Table~\ref{tab:bootstrap_fit_summary}.
Each colored distribution in Figure~\ref{fig:bootstrap_histogram} corresponds to the set of simulated half-life values for one of the original measurements: \(418 \pm 15\)~years (HL3\_1), \(437.7 \pm 7.7\)~years (HL3\_2), and \(448 \pm 27\)~years (HL3\_3).
For each dataset, 100,000 bootstrap replicates were generated by randomly sampling according to the reported measurement uncertainties, assuming Gaussian distributions.
The grey histogram in the background shows the combined distribution of all simulated values.
The plotted histograms illustrate the spread and central tendency of the simulated half-lives, with clear separation between the peaks corresponding to each original measurement.

Table~\ref{tab:bootstrap_fit_summary} provides a quantitative comparison between the fitted Gaussian parameters extracted from the simulated datasets and the original measured values. 
For each origin, the mean (\(\mu\)) and standard deviation (\(\sigma\)) obtained from the bootstrap simulations are compared to the original published values. 
The absolute percent differences (\(\Delta \mu\) and \(\Delta \sigma\)) are extremely small, all below \(0.5\%\), indicating excellent agreement between the simulations and the experimental data. 
This validates that the bootstrap sampling process accurately preserved the statistical properties of the original measurements without introducing additional bias or systematic deviation.

The close agreement between the MFV values derived from the five-element and three-element datasets demonstrates that the central estimate remains stable even when early, high-uncertainty or outlier data are excluded. It further underscores the MFV-HPB method’s ability to yield reliable central values from small, heterogeneous, and historically diverse datasets. This reinforces its usefulness in validating or reinterpreting legacy nuclear measurements using modern statistical approaches.

In both cases, the confidence intervals were estimated using the percentile method, which extracts quantiles from the empirical bootstrap distribution. To improve statistical accuracy and reduce variability, the MFV was computed for each of 100{,}000 bootstrap samples. This large number of resamples ensures precise estimation of the confidence bounds even with limited source data and by considering the uncertainty associated with each element in the dataset (see Table~\ref{tab:bootstrap_fit_summary}). In addition, a large bootstrap sample size was selected to minimize the introduction of additional statistical errors due to the inherent randomness of the HPB technique. To validate this, we performed a Gaussian fit for each individual data point generated during resampling and confirmed that both the mean and standard deviation of the fitted distribution matched the original source values. This step confirmed that the HPB procedure truly preserved the original data structure and uncertainty, thereby ensuring the reliability of the resulting MFV-based confidence intervals.

An important observation from this analysis is that the confidence intervals derived using the MFV-HPB method are not necessarily symmetric around the central value. This contrasts with intervals based on the arithmetic mean, which typically assume normally distributed data and produce symmetric intervals. The asymmetry observed in the MFV-HPB results more accurately reflects the underlying distribution of the data and highlights the method's suitability for non-Gaussian datasets.

Despite its strengths, the MFV-HPB framework is not without limitations. Like other statistical methods, it performs best when most data are concentrated near the true value. If erroneous or biased values overwhelm a dataset, even the MFV-HPB approach can yield misleading results. This limitation is not unique to MFV; it represents a general challenge in statistical inference that underscores the importance of high-quality data.

In addition, the HPB method is computationally intensive, especially when large numbers of bootstrap samples are required or when applied to larger datasets. Nonetheless, this approach remains highly effective in situations in which measurement uncertainty is significant and traditional statistical assumptions do not hold. The flexibility and reliability of the proposed method make it a valuable tool not only in nuclear data analysis but also in other fields, such as biophysics and medical diagnostics, where datasets are often small, noisy, and irregularly distributed.

\hl{The MFV-HPB framework is a method developed to provide strong estimates of central values and their confidence intervals in datasets with variability and potential outliers. This framework faces specific challenges when dealing with datasets with multimodal distributions, which are common in areas like molecular dynamics and microscale mechanical mapping of cancer cells, where bimodal or multimodal systems often occur}~\cite{NIKOLIC20223586,wuIdentificationNovelTargets2011a}. \hl{The MFV-HPB procedure can be performed in a ``mode-by-mode'' approach. By isolating each peak and recalculating the MFV with a confidence interval using a hybrid-bootstrap method, this approach usually offers more precise confidence intervals for each mode. The MFV step reduces the impact of extreme data points within each subgroup, leading to MFV-HPB confidence intervals that are systematically narrower than those from traditional methods, thus providing more insightful summaries of multimodal datasets.}

\hl{In this study, the dataset represents a single physical parameter: the cross-section value of a particular nuclear reaction at a specific neutron energy. Ideally, the data should converge to a value that accurately reflects this parameter. However, multiple peaks or modes can appear in the data. These multimodal features often result from differences in methods, inconsistent measurement techniques, or hidden systematic errors rather than actual physical variations.}

\hl{In the analysis of neutron lifetime data}~\cite{zhang2022mfv}, \hl{different experimental methods produced several groups of values. Although these groups varied, there should be one true central value for the neutron lifetime because it is a fundamental physical constant. The MFV approach helped find a reliable central value that considered all data points and reduced the effects of clusters caused by specific measurement techniques.}

\hl{Similarly, this study initially observed a bimodal pattern in the cross-section data for the reaction $^{109}\text{Ag}(\text{n},2\text{n})^{108\text{m}}\text{Ag}$ (see Figure}~\ref{fig:histogram}). \hl{This pattern mainly arises from differences in the reported half-life of $^{108\text{m}}\text{Ag}$. Some values relied on older half-life estimates, whereas others used updated measurements. Adjusting the dataset to correct these systematic differences reduced the bimodal effect (see Figure}~\ref{fig:histogram1}), \hl{supporting the conclusion that a single value can accurately describe the cross-section.}

\hl{The MFV-HPB framework considers the uncertainties of each data point, thereby providing a more complete estimate of the central value and its confidence intervals. In contrast, PDG uses a weighted mean method, which also considers uncertainties but tends to bias the central value toward data points with the smallest uncertainties}~\cite{golovko2025hpb_IS}. \hl{This bias can become an issue when the smallest uncertainty does not truly reflect better measurement accuracy but instead arises from unrecognized systematic errors or possible mistakes in data reporting}~\cite{golovko2023Ar39}.

\section{ Conclusion }

In this study, we presented a robust statistical approach that combines Steiner's most frequent value  method with a hybrid bootstrap procedure  to estimate central tendency and confidence intervals for datasets affected by variability, outliers, or limited sample size. We applied this method to fast-neutron activation cross-section measurements of the \(^{109}\text{Ag}(\text{n},2\text{n})^{108\text{m}}\text{Ag}\) reaction at \(14.7 \pm 0.2~\mathrm{MeV}\), a dataset characterized by significant uncertainties and inconsistencies.

Our findings show that the MFV provides a stable and reliable central estimate, even when traditional metrics like the arithmetic or weighted mean are distorted by outliers. Combined with the hybrid bootstrap procedure, which models both measurement uncertainty and sampling variability, the method produced interpretable and well-constrained confidence intervals. For the cross-section data, the MFV was determined to be \(709~\mathrm{mb}\), with a 68.27\% confidence interval of \([691, 744]~\mathrm{mb}\) and a 95.45\% confidence interval of \([666, 774]~\mathrm{mb}\), based on 500,000 hybrid bootstrap replicates.

To further demonstrate the method’s flexibility, we applied the MFV-HPB framework to estimate the half-life of the \(^{108\text{m}}\text{Ag}\) isotope using a small dataset of published measurements. When analyzing five available values (excluding the earliest lower-limit estimate from 1960), the MFV was found to be \(433.5~\mathrm{years}\), with a 68.27\% confidence interval of \([300.6, 441.1]~\mathrm{years}\) and a 95.45\% interval of \([124.4, 456.6]~\mathrm{years}\). A secondary analysis, considering only the three most recent and precise measurements, yielded an MFV of \(439.1~\mathrm{years}\) with narrower confidence intervals. The close agreement between these results confirms that the method maintains stability even when earlier, higher-uncertainty data are excluded.

Importantly, the MFV-HPB method does not assume any particular shape for the underlying data distribution. As observed in the half-life analysis, the resulting confidence intervals can be asymmetric, providing a more accurate reflection of the real uncertainty structure compared to traditional Gaussian-based methods.

The hybrid bootstrap method used in this study combines non-parametric resampling of the original data entries with parametric simulation based on their individual uncertainties. Unlike a purely parametric bootstrap, where each original data point contributes exactly one simulated value per replicate, our approach first resamples the data points with replacement. This adds an extra layer of variability, capturing uncertainty about which measurements are most representative. Especially in small datasets or when uncertainties are heterogeneous--as in early half-life measurements--this strategy produces slightly wider and more conservative confidence intervals, better representing the true uncertainty.

Although this work focused on nuclear physics examples, the statistical challenges addressed here--such as data sparsity, measurement uncertainty, and non-Gaussian behavior--are common across many fields, including biomolecular research, environmental monitoring, and diagnostics. As a result, the MFV-HPB framework provides a generalizable and reliable tool for extracting trustworthy insights from real-world data.

We hope that this study will encourage broader adoption of the MFV-HPB methodology in scientific data analysis, particularly in situations where robustness and interpretability are critical. Future efforts could expand its application to additional scientific domains and help formalize its integration into standard evaluation practices. Ultimately, the MFV-HPB approach offers a statistically rigorous and adaptable pathway for addressing uncertainty in complex datasets.

\authorcontributions{
The author performed the conceptualization, methodology, formal analysis, investigation, original draft preparation, review and editing, and visualization. The author has read and approved the final version of the manuscript for publication.
}

\funding{This research received no external funding.}

\institutionalreview{Not applicable.}

\dataavailability{This manuscript has associated data are available in the following repository: \href{https://osf.io/g2h3m/}{https://osf.io/g2h3m/} \cite{golovkoSupportingDataset2025}. To validate the reliability of the method, the author used published neutron lifetime data as a benchmark (included in the repository). This confirmed that the MFV estimate was consistent with the findings of the original study~\cite{zhang2022mfv}.}

\acknowledgments{
I would like to sincerely thank Maria Filimonova for her invaluable support and assistance throughout this work. I am also grateful to the management and staff at Canadian Nuclear Laboratories for providing a supportive environment for this study, with special thanks to Genevieve Hamilton and David Yuke. I also appreciate the thoughtful comments and suggestions from the anonymous reviewers, which have helped improve the quality of this paper.

\textbf{Declaration of generative AI and AI-assisted technologies: } During the preparation of this work, the author used ChatGPT to check the language of the manuscript. After using this tool, the author reviewed and edited the content as needed and takes full responsibility for the content of the publication.
}

\conflictsofinterest{The authors declare  that they have no conflicts of interest.}

\abbreviations{Abbreviations}{
The following abbreviations were used in this manuscript:\\

\noindent
\begin{tabular}{@{}ll}
	CI   & Confidence interval \\
	CRL  & Chalk River Laboratories \\
	HPB  & Hybrid parametric bootstrap \\
	MeV  & Mega electron volt \\
	MFV  & Most frequent value \\
	KL   & Kullback--Leibler \\
\end{tabular}
}

\begin{adjustwidth}{-\extralength}{0cm}

\reftitle{References}

\end{adjustwidth}
\end{document}